\documentclass[aps,prl,twocolumn,noshowpacs]{revtex4-1}

\usepackage{bbm}
\usepackage{mathrsfs}
\usepackage{amsmath}
\usepackage{amsfonts}
\usepackage[colorlinks=true,citecolor=blue,anchorcolor=blue]{hyperref}
\usepackage{graphicx,epstopdf}
\usepackage{subfigure}
\usepackage{epsfig}
\usepackage{dcolumn}
\usepackage{bm}
\usepackage{color}
\usepackage{xcolor}

\begin{document}
\title{Dynamical axion state with hidden pseudospin Chern numbers in MnBi$_{2}$Te$_{4}$-based heterostructures}

\author{Huaiqiang Wang$^{1,\dagger}$, Dinghui Wang$^{1,\dagger}$, Zhilong Yang$^{1}$, Minji Shi$^1$, Jiawei Ruan$^1$, Dingyu Xing$^{1,2}$, Jing Wang$^{3,2,4\ast}$, Haijun Zhang$^{1,2\ast}$}

\affiliation{
$^1$ National Laboratory of Solid State Microstructures, School of Physics, Nanjing University, Nanjing 210093, China\\
$^2$ Collaborative Innovation Center of Advanced Microstructures, Nanjing University, Nanjing 210093, China\\
$^3$ State Key Laboratory of Surface Physics, Department of Physics, Fudan University, Shanghai 200433, China\\
$^4$ Institute for Nanoelectronic Devices and Quantum Computing, Fudan University, Shanghai 200433, China
}

\email{zhanghj@nju.edu.cn;\\ wjingphys@fudan.edu.cn}

\begin{abstract}
Axion is a hypothetical elementary particle which was initially postulated to solve the charge conjugation-parity problem in particle physics. Interestingly, the axion state has emerged in effective theory of topological insulators and has attracted extensive attention in condensed matter physics. Time-reversal or inversion symmetry constrains the axion field $\theta$ to be quantized. When both the time-reversal and inversion symmetries are broken by, say, an antiferromagnetic order, the axion field $\theta$ could become unquantized and dynamical along with magnetic fluctuations, which is termed the dynamical axion field. Here, we reveal that a wide class of topological-insulator-based dynamical axion states could be distinguished from the normal-insulator-based ones by a hidden quantity derived from the pseudospin Chern number. Motivated by recent research on MnBi$_{2}$Te$_{4}$-family materials, we further show that such topological-insulator-based dynamical axion states can be hopefully achieved in MnBi$_{2}$Te$_{4}$-based heterostructures, which should greatly facilitate the study of axion electrodynamics in condensed matter physics.

\end{abstract}
\maketitle

\emph{Introduction.} Recently, many analogs of elementary particles, which were originally discovered in particle physics, have been found as quasi-particle excitations in condensed matter physics, such as, for example, the massless Weyl and Dirac fermions in topological semimetals \cite{Armitage2018Weyl} and mysterious Majorana fermions in topological superconductors\cite{Qi2011, Elliott2015}. Among them, the search for an axion state has attracted intensive interest in condensed matter physics~\cite{Qi2008, Essin2009, Rosenberg2010, Karch2009, Bermudez2010, Nomura2011, Wan2012, Sekine2014, Wang2015Quantized, Morimoto2015, Turner2012, Wang2013Chiral, Goswami2014, Lee2015, You2016, wieder2018axion, yue2019symmetry, xu2019higher, hou2019Baxion,otrokov2019unique, zhang2018topological, li2019intrinsic, chowdhury2019prediction, Wu2016, Grauer2017, Mogieaao1669, Okada2016Terahertz, Mogi2017A, Dziom2017Observation, Xiao2018, Varnava2018, gong2019experimental, chen2019searching, otrokov2018prediction, gui2019new, Allen2019, Chen2019Supp, gooth2019evidence}. As a hypothetical elementary particle, the axion is a possible component of dark matter of the Universe~\cite{Pecci1977}. Axions gained new vigor from the effective theory of topological insulators (TIs)~\cite{Qi2008, Qi2011}, which is described by an axion Lagrangian density $\mathcal{L}_{\theta}=(\theta/2\pi)(\alpha/2\pi)\mathbf{E}\cdot \mathbf{B}$. Here, $\alpha=e^{2}/\hbar c$ is the fine structure constant, $\mathbf{E}$ and $\mathbf{B}$ represent electric and magnetic fields, and the dimensionless pseudoscalar $\theta$ is a parameter describing the insulator, which is referred to as the axion field in axion electrodynamics~\cite{Wilczek1987}. Generically, $\theta$ is odd under time-reversal symmetry (TRS) and inversion symmetry (IS). With periodic boundary conditions in space-time, $\theta$ has a $2\pi$-periodicity and it is constrained to $\pi$ for time-reversal-invariant TIs and $0$ for normal insulators (NIs). Such quantized $\theta$ in TIs will lead to the topological magnetoelectric effect~\cite{Qi2008, Essin2009, Rosenberg2010, Karch2009, Bermudez2010, Vazifeh2010, Nomura2011, Sekine2014, Wang2015Quantized, Morimoto2015} and unique magneto-optical Faraday and Kerr effects~\cite{Tse2010, Maciejko2010, Ochiai2012Theory, Mal2013, Wu2016}.

Once both TRS and IS are broken, $\theta$ deviates from the quantized value of $0$ or $\pi$, as has been shown for TIs in the presence of antiferromagnetic (AFM) order~\cite{li2010dynamical, Wang2011Topological} and antiferromagnet chromia~\cite{hehl2009magnetoelectric}. Intriguingly, an inherently fluctuating AFM order will induce fluctuations of the axion field $\theta$ as a dynamical variable with spatial and temporal dependence~\cite{li2010dynamical, Wang2011Topological, Sekine2014, Wang2016Dynamical, wang2019dynamic}. Such a dynamical axion field could give rise to rich, interesting effects, such as, the dynamical chiral magnetic effect and anomalous Hall effect~\cite{Wilczek1987, Sekine2016Chiral, Zhang2019Dynamical}, instability in an external electric field~\cite{Ooguri2012Insta, imaeda2019axion}, and axionic polaritons~\cite{li2010dynamical}.

Up to now, two important issues on the dynamical axion state still remain open. First, since the broken TRS invalidates the $Z_{2}$ topological invariant in TIs~\cite{fu2007topological}, it remains unknown whether it is available to distinguish the TI-based dynamical axion state ~\cite{li2010dynamical} with a large static $\theta\sim\pi$ from the NI-based one with a small static $\theta\sim 0$~\cite{Coh2011}. Second, the experimental demonstration of a TI-based dynamical axion state is significantly hindered by the lack of realistic materials. Here, we explicitly address the above two issues in the affirmative. First, based on an effective model analysis, we uncover a hidden quantity derived from pseudospin Chern numbers~\cite{Sheng2006Quantum, Prodan2009, Yang2011} to differentiate a wide class of TI- and NI-based dynamical axion states. Further, inspired by recent studies on the MnBi$_{2}$Te$_{4}$ family of intrinsic magnetic insulators~\cite{zhang2018topological, li2019intrinsic, gong2019experimental, otrokov2018prediction, vidal2019massive, chen2019searching, deng2019magnetic, liu2019quantum, zhang2019experimental, sun2019rational, Vidal2019}, we find that TI-based dynamical axion states can be hopefully achieved in MnBi$_{2}$Te$_{4}$$/$Bi$_{2}$Te$_{3}$-based magnetic van der Waals (vdW) heterostructures through breaking both TRS and IS by specific structural designs. Interestingly, some natural vdW heterostructures of (MnBi$_{2}$Te$_{4}$)$_{m}$(Bi$_{2}$Te$_{3}$)$_n$ have been experimentally fabricated~\cite{hu2019van, wu2019natural, Vidal2019}, which may provide an ideal platform to study the axion electrodynamics and to detect the axion dark matter~\cite{Marsh2019}.\\

\emph{Low-energy effective model.} Without loss of generality, we start from the typical low-energy effective model of Bi$_{2}$Te$_{3}$-family TIs~\cite{Zhang2009Topological} and then treat the AFM order as a perturbation~\cite{li2010dynamical}. In the nonmagnetic state with the preserved IS, the four low-lying states are the bonding states $|P1^{+}_{z},\uparrow(\downarrow)\rangle$ from Bi atoms and antibonding states $|P2^{-}_{z},\uparrow(\downarrow)\rangle$ from Te atoms, where the superscript ``$+$'' (``$-$'') represents positive (negative) parity. In the ordered basis of ($|P1^{+}_{z},\uparrow\rangle$, $|P1^{+}_{z},\downarrow\rangle$, $|P2^{-}_{z},\uparrow\rangle$, $|P2^{-}_{z},\downarrow\rangle$), the symmetry operations of the $D_{3d}$ group at $\Gamma$ are represented, for example, as TRS $\mathcal{T}=\tau_{0}\otimes i\sigma_{y}K$, IS $\mathcal{P}=\tau_{z}\otimes \sigma_{0}$, rotation symmetries $C_{3z}=\mathrm{exp}[\tau_{0}\otimes (-i\pi/3)\sigma_{z}]$ and $C_{2x}=\mathrm{exp}[\tau_{0}\otimes (-i\pi/2)\sigma_{x}]$, where $K$ denotes the complex conjugate, $\tau_{0}$ and $\sigma_{0}$ are $2\times 2$ identity matrices, and $\tau_{x,y,z}$ and $\sigma_{x,y,z}$ are Pauli matrices acting in the orbital and spin spaces, respectively. Keeping the most relevant terms up to quadratic order of $k$, the effective four-band Hamiltonian is given by
\begin{equation}
\label{dkgamma}
\begin{split}
\mathcal{H}_{0}(\mathbf{k})=&\epsilon_{0}(\mathbf{k})\mathbb{I}_{4\times 4}+\sum^{5}_{i=1}d_{i}(k)\Gamma^{i},\\
d_{1,2,...,5}(\mathbf{k})=&\Big(A_{2}k_{y},-A_{2}k_{x},A_{1}k_{z},M(\mathbf{k}),0 \Big),
\end{split}
\end{equation}
where $\epsilon_{0}(\mathbf{k})=C+D_{1}k_{z}^{2}+D_{2}(k_{x}^{2}+k_{y}^{2})$, $M(\mathbf{k})=M+B_{1}k_{z}^{2}+B_{2}(k_{x}^{2}+k_{y}^{2})$, and the five Dirac matrices are represented as $\Gamma^{1,2,...,5}=(\tau_{x}\otimes\sigma_{x},\tau_{x}\otimes\sigma_{y},\tau_{y}\otimes \sigma_{0}, \tau_{z}\otimes \sigma_{0},\tau_{x}\otimes\sigma_{z})$, which satisfy the Clifford algebra $\{\Gamma^{i},\Gamma^{j}\}=2\delta_{i,j}$. The TI phase requires the condition of $M/B_{i=1,2}<0$. Hereafter, we assume $B_{i=1,2}$ to be positive, and $M<0$ ($M>0$) corresponds to the TI (NI).

The AFM order breaks both $\mathcal{P}$ and $\mathcal{T}$ simultaneously, which is essential for the emergence of dynamical axion states, and to the leading order, it will give rise to the $\mathcal{P}$-, and $\mathcal{T}$- breaking mass term $\delta\mathcal{\mathcal{H}}_{\mathrm{AFM}}=m_{5}\Gamma_{5}$~\cite{li2010dynamical}. Furthermore, if the AFM order preserves the combined $\mathcal{PT}$ symmetry, only the identity matrix $\mathbb{I}_{4\times 4}$ and the above five linearly independent anticommuting Dirac matrices $\Gamma^{1,2,...,5}$ are allowed~\cite{Supp}. Consequently, the total Hamiltonian is written as $\mathcal{H}(\mathbf{k})=\mathcal{H}_{0}(\mathbf{k})+m_{5}\Gamma_{5}$, and the antiunitary $\mathcal{PT}$ symmetry with $\mathcal{(PT)}^{2}=-1$ ensures the double degeneracy of each band. It should be emphasized that due to the nonvanishing AFM mass term, when tuning $M$ from negative (TI side) to positive (NI side), there will be no topological phase transition, but a crossover with an avoided gap closing. Nevertheless, the two magnetic states in the TI and NI sides, respectively, can still be differentiated by a hidden $Z_{2}$-like quantity derived from the pseudospin Chern numbers as we show below.\\
\begin{figure}
  \centering
  \includegraphics[width=3in]{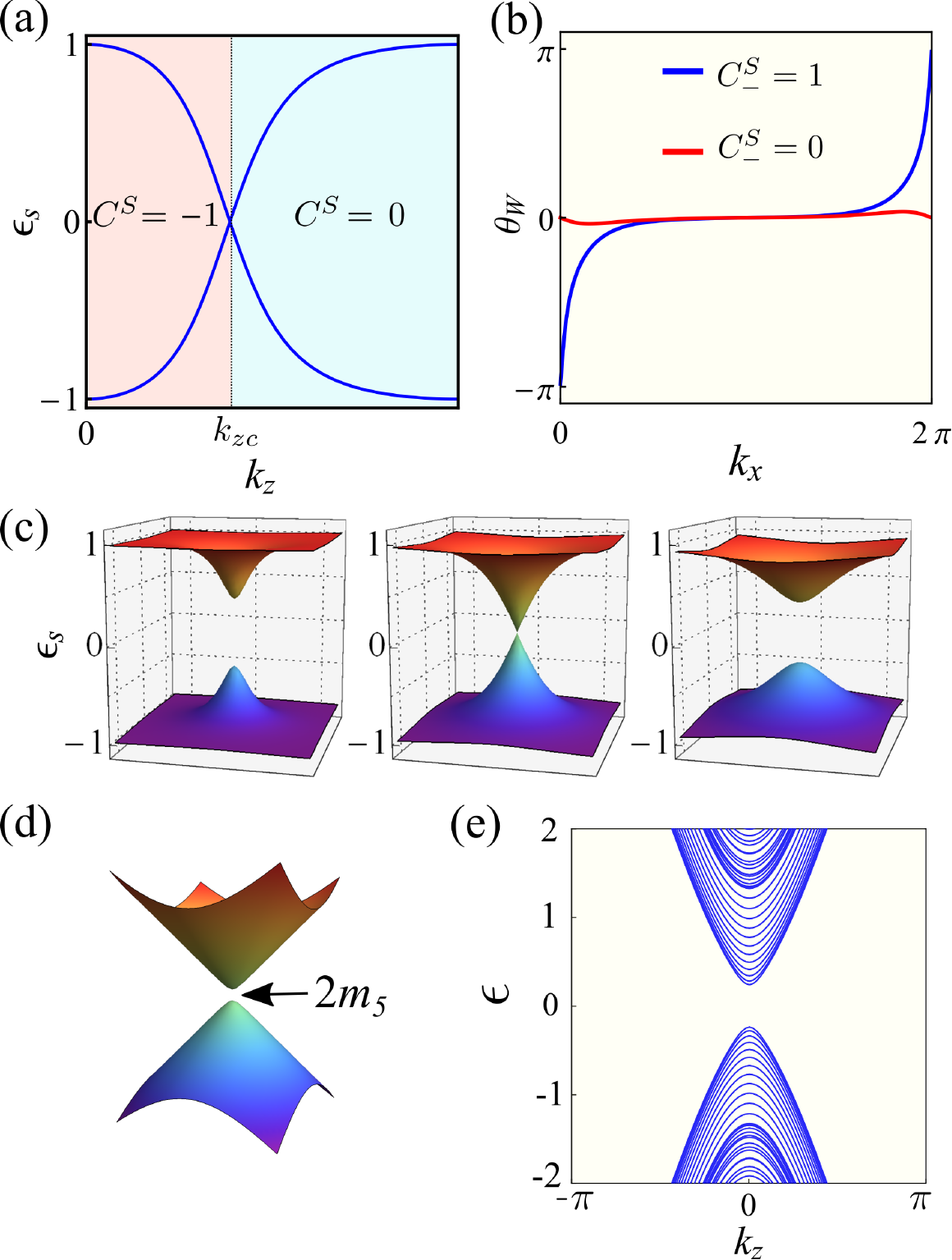}\\
  \caption{(a) The $k_{z}$-dependent pseudospin spectrum at $k_{x}=k_{y}=0$, with the gap closing at $k_{zc}=\sqrt{-M/B_{1}}$. (c) Three representative pseudospin spectra in the 2D slice ($k_{x}$-$k_{y}$) for $|k_{z}|<k_{zc}$ (left), $|k_{z}|=k_{zc}$ (middle), and $|k_{z}|>k_{zc}$ (right). (b) Representative $U(1)$ Wilson loops for the lower branch of the pseudospin spectra in the 2D slice ($k_{x}$-$k_{y}$) of a lattice model as a function of $k_{x}$ when $C_{-}^{S}(k_{z})=1$ (blue line) and 0 (red line), respectively. (d) Schematic illustration of the massive Dirac surface state with a $2m_5$ gap inherited from the bulk bands. (e) Energy spectrum along the $k_z$ direction with open-boundary conditions in both $x$ and $y$ directions. The parameters are typically chosen as $C=D_{1}=D_{2}=0$, $A_{1}=A_{2}=B_{1}=B_{2}=1$, $M=-0.3$, and $m_{5}=0.03$, which are given in dimensionless form for simplicity.}\label{fig1}
\end{figure}

\emph{Pseudospin Chern number.} The (pseudo)spin Chern number was originally proposed in the quantum spin Hall effect~\cite{Sheng2006Quantum}, which remains valid and robust even when TRS is broken or the (pseudo)spin is not conserved~\cite{Sheng2006Quantum, Prodan2009, Yang2011, Li2010Chern}.  The definition of the pseudospin Chern number relies on a smooth decomposition of the occupied valence bands into two pseudospin sectors through projecting a pseudospin operator $\hat{S}$ into valence bands as $h_S=P_{v}\hat{S}P_{v}$ with $P_{v}$ as the valence-band projecting operator~\cite{Sheng2006Quantum, Prodan2009, Yang2011, Supp}. Diagonalization of $h_S$ then gives rise to two pseudospin branches with eigenvalues $\pm\epsilon_{S}(\mathbf{k})$, and corresponding eigenvectors $|\varphi_{\pm}\rangle$. For each pseudospin branch, the usual Chern number can be calculated in two-dimensional (2D) momentum space as
\begin{equation}
\label{spin Chern}
C_{\pm}^{S}=\frac{1}{2\pi}\int d^{2}k\Omega_{\pm}^{\mu\nu},
\end{equation}
where $\Omega_{\pm}^{\mu\nu}=-2\mathrm{Im}\langle\frac{\partial\varphi_{\pm}}{\partial k_{\mu}}| \frac{\partial\varphi_{\pm}}{\partial k_{\nu}}\rangle$ is the Berry curvature in the 2D $k_{\mu}$-$k_{\nu}$ slice. The pseudospin Chern number is then defined as $C^{S}=(C_{+}^{S}-C_{-}^{S})/2$.

By treating one momentum as a parameter, we can calculate $C^{S}$ for each 2D slice in the 3D Brillouin zone. Specifically, we treat $k_{z}$ as the parameter and investigate the 2D slices ($k_{x}$-$k_{y}$). We construct the pseudospin operator as $\hat{S}=\tau_{z}\sigma_{z}$ in the current basis of $\mathcal{H}(\mathbf{k})$~\cite{Supp}. The $k_{z}$-dependent pseudospin Chern number can then be analytically derived as~\cite{Supp}
\begin{equation}
\label{Ckz}
C^{S}(k_{z})=\frac{1}{2}\Big[-\mathrm{sgn}(B_{2})+\mathrm{sgn}(M+B_{1}k_{z}^{2})\Big].
\end{equation}
Interestingly, it is independent of the AFM mass $m_{5}$. Since $B_{1}$ and $B_{2}$ are positive, $C^{S}(k_{z})$ always equals zero in the NI-based regime with $M>0$, whereas one can obtain $C^{S}(k_{z})=-1$ for $|k_{z}|<k_{zc}=\sqrt{-M/B_{1}}$ and $C^{S}(k_{z})=0$ for $|k_{z}|>k_{zc}$ in the TI-based regime with $M<0$, as shown in Fig.~1(a). It is noteworthy that the change of $C^{S}(k_{z})$ here is accompanied by the closing of the pseudospin-spectrum gap instead of the energy gap which is always kept open. This can be seen by the $k_{z}$-dependent pseudospin spectrum $\epsilon_{S}$ at (0,0,$k_{z}$) in Fig.~1(a) with a closing point at $|k_{z}|=k_{zc}$. Representative pseudospin spectra in the $k_{x}$-$k_{y}$ plane are also shown in Fig.~1(c) for $|k_{z}|<k_{zc}$ (left column), $|k_{z}|=k_{zc}$ (middle column), and $|k_{z}|>k_{zc}$ (right column), respectively.
Moreover, we stress that the pseudospin Chern number remains quantized even though the pseudospin is not conserved for $k_{z}\neq 0$ ($\epsilon_{S}$ deviates from $\pm1$)~\cite{Prodan2009}. This can alternatively be reflected by the $U(1)$ Wilson loop spectra~\cite{Yu2011Equivalent} for each pseudospin branch in the 2D slice ($k_{x}$-$k_{y}$) of a lattice model~\cite{Supp}. As shown in Fig.~1(b), when $k_x$ goes from $0$ to $2\pi$, the Wilson loop of the lower pseudospin branch exhibits one (no) winding for $C_{-}^{S}(k_z)=+1 (0)$.

In the TI-based magnetic state, such a change of $C^{S}$ from $k_{z}=0$ to $|k_{z}|>k_{zc}$ essentially results from the inverted band structure ($M<0$) at the $\Gamma$ point of the parent TI. This enables us to define a $Z_{2}$-like quantity $\nu$ as
\begin{equation}
\label{z2}
\nu=(-1)^{C^{S}(k_{z}=0)+C^{S}(k_{z}=\infty)},
\end{equation}
where $k_{z}=\infty$ should be replaced by $k_{z}=\pi$ in Bloch bands, and $\nu=-1$ (or $+1$) corresponds to the TI-based (NI-based) state. It is worth mentioning that when treating $k_{x}$ (or $k_{y}$) as a parameter, through similar procedures of calculating $k_{x}$- (or $k_{y}$-) dependent pseudospin Chern numbers~\cite{Supp}, identical $\nu$ can be obtained. Strictly speaking, $\nu$ is not a topological invariant since the above two magnetic states are topologically the same with an avoided gap-closing crossover. Nonetheless, a ``nontrivial'' $\nu=-1$ still acts as a useful quantity which captures the band-inversion information of the parent TI state. Such a quantity can be applied to a wide class of 3D magnetic insulators, as long as the time-reversal-invariant 2D plane could be effectively reduced to a Bernevig-Hughes-Zhang-like model~\cite{bernevig2006quantum, Bernevig2013Topological} by some unitary transformations without the TRS-breaking magnetic order.\\

\emph{Massive Dirac surface states.} For a 3D TI, gapless surface states, with a single massless Dirac cone, emerge on the surface. However, due to the AFM mass term in TI-based states, surface states are gapped with a single massive Dirac cone on all surfaces, which can be clearly seen by transforming the above continuum model to a tight-binding one on a cubic lattice. The massive Dirac-cone surface states along the $z$ direction can be derived as $h^{z}_{\mathrm{surf}}=\pm\big[A_{2}(k_{y}\sigma_{x}-k_{x}\sigma_{y})+m_{5}\sigma_{z}\big]$~\cite{Supp}, where ``$+$'' and ``$-$'' refer to the bottom and top surfaces, respectively, indicating an energy gap of $2m_{5}$ for the surface bands, as shown in Fig. 1(d). Similarly, for the surface states along $x$ and $y$ directions, the effective Hamiltonians are given by $h^{x}_{\mathrm{surf}}=\mp(A_{2}k_{y}\sigma_{z}+A_{1}k_{z}\sigma_{y}+m_{5}\sigma_{x})$ and $h^{y}_{\mathrm{surf}}=\pm(-A_{2}k_{x}\sigma_{z}+A_{1}k_{z}\sigma_{x}-m_{5}\sigma_{y})$, respectively~\cite{Supp}.
Note that all surfaces are gapped for our model and that no one-dimensional gapless hinge states are found, as shown in Fig. 1(e) by the energy spectrum with open-boundary conditions in both the $x$ and $y$ directions. This is different from the recently proposed (static) axion insulators with chiral hinge states between neighboring gapped surface states~\cite{yue2019symmetry, xu2019higher}, due to the lack of IS and higher-order topology in our system.
\\

\begin{figure}[t]
  \centering
  \includegraphics[width=3in]{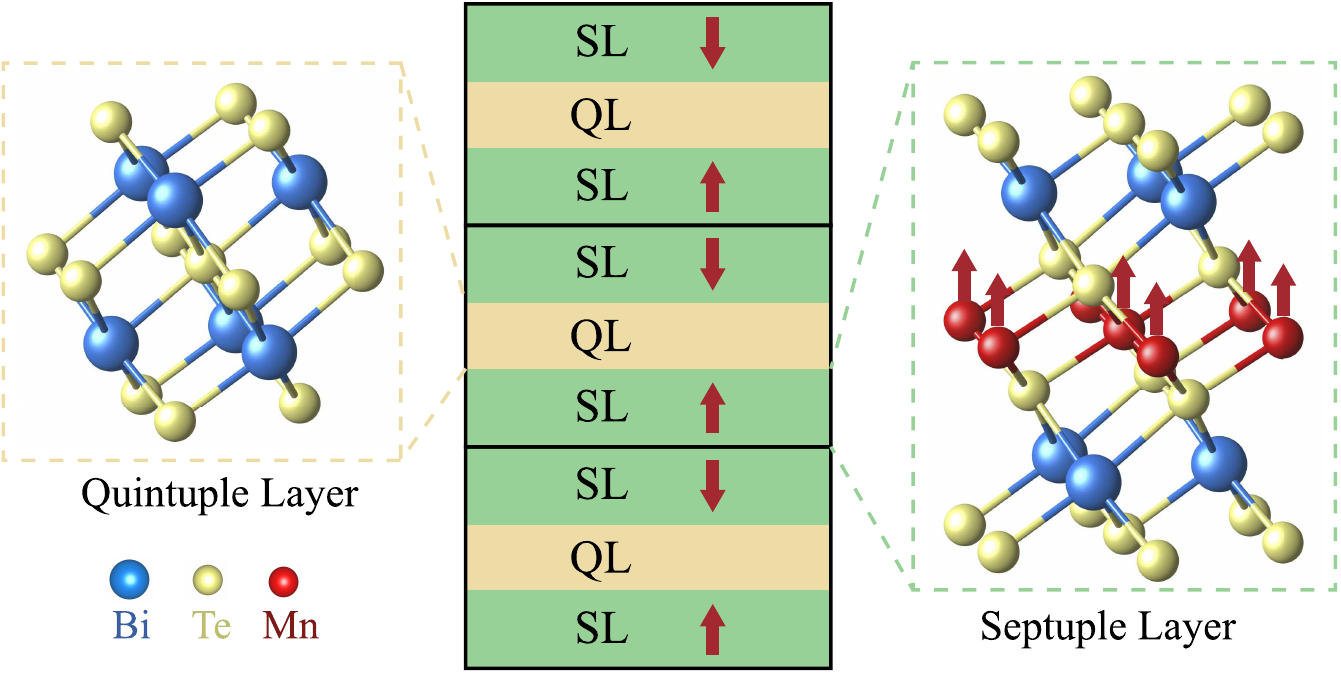}\\
  \caption{Schematic illustration of the crystal structures of the AFM (MnBi$_{2}$Te$_{4}$)$_{2}$(Bi$_{2}$Te$_3$) superlattice, where the unit cell consists of a Bi$_{2}$Te$_3$ quintuple layer (QL) sandwiched between two MnBi$_{2}$Te$_{4}$ septuple layers (SLs).}\label{fig2}
\end{figure}

\emph {Magnetic van der Waals heterostructures.} The recently discovered MnBi$_{2}$Te$_{4}$-family insulators host the static axion state with $\theta=\pi$, because the IS is preserved~\cite{zhang2018topological}. It is promising to design (MnBi$_{2}$Te$_{4}$)$_{m}$-(Bi$_{2}$Te$_{3}$)$_n$ superlattices, where both $\mathcal{T}$ and  $\mathcal{P}$ are broken with suitably chosen $m$ and $n$, to realize the TI-based dynamical axion state. Here, we take (MnBi$_{2}$Te$_{4}$)$_{2}$(Bi$_{2}$Te$_{3}$) as a concrete example, also written as Mn$_{2}$Bi$_{6}$Te$_{11}$.  Its crystal structure is schematically shown in Fig.~2. The basic stacking block consists of one Bi$_{2}$Te$_3$ quintuple layer (QL) sandwiched between two MnBi$_{2}$Te$_{4}$ septuple layers (SLs). Based on first-principles calculations, the magnetic structure is the $A$-type AFM order between nearest SLs and the out-of-plane ferromagnetic order in each SL. The exchange interaction in Mn$_{2}$Bi$_{6}$Te$_{11}$ is expected to be weaker than that in MnBi$_{2}$Te$_{4}$ due to the intercalation layer Bi$_{2}$Te$_{3}$.

\begin{figure}
  \centering
  \includegraphics[width=3.4in]{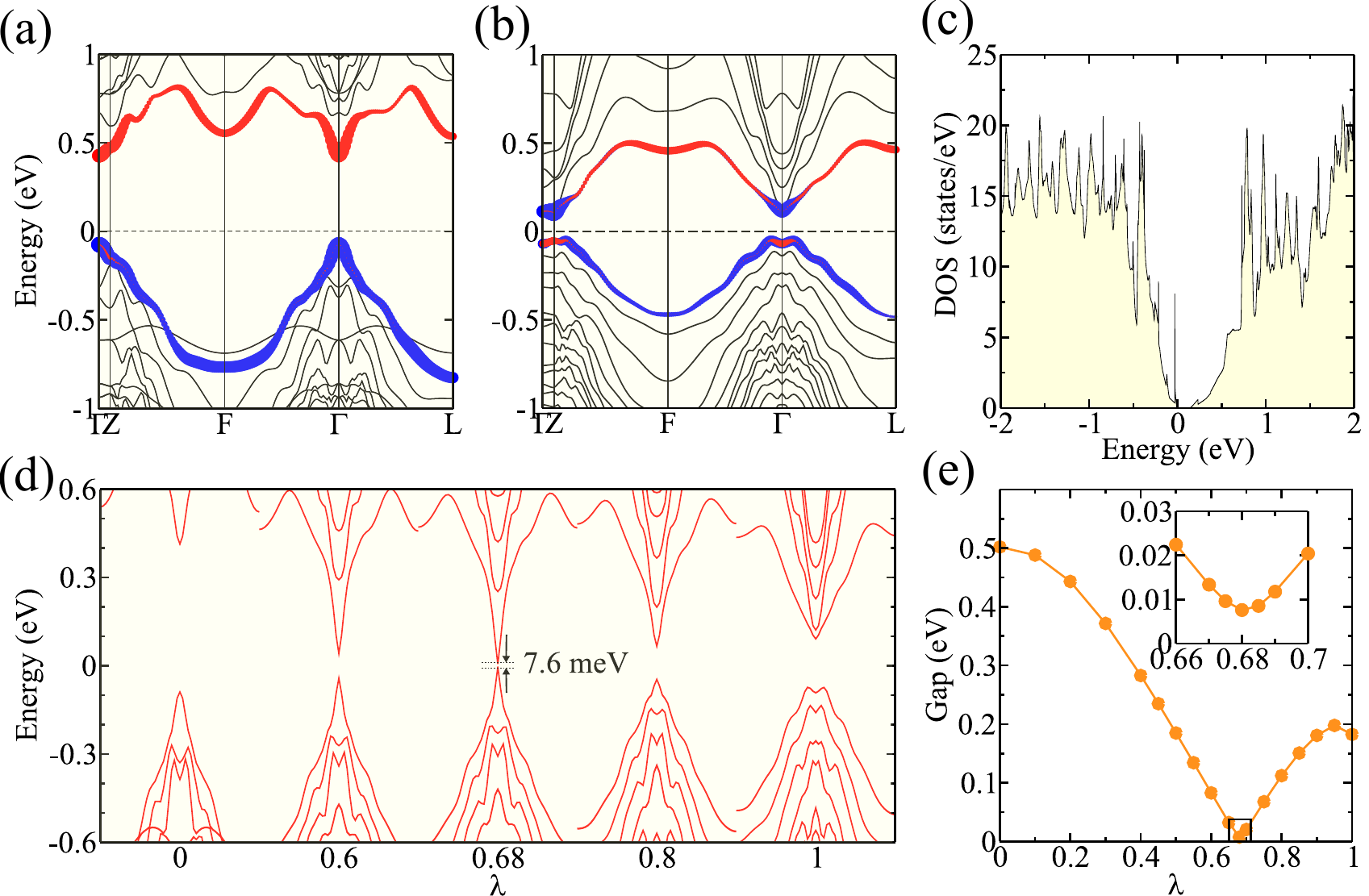}\\
  \caption{(a), (b), The band structures (a) without and (b) with SOC. The fat bands indicate a SOC-induced ``band inversion'' between $p$ orbitals of Bi (red) and Te (blue) around the Fermi energy. (c) The  density of states near the Fermi energy. (d) The evolution of the band structure with increasing the SOC strength $\lambda\ast\lambda_0$, where $\lambda_0$ denotes the realistic SOC strength. (e) The dependence of the band gap on the SOC strength. Inset: The minimum gap ($\sim7.6$ meV) at about $\lambda=0.68$.}\label{fig3}
\end{figure}

If ignoring the magnetism, Mn$_{2}$Bi$_{6}$Te$_{11}$ takes the rhombohedral crystal structure with the same space group $D_{3d}^{5}$ ($R\bar{3}m$) as Bi$_{2}$Te$_{3}$. Each atomic layer has a triangular lattice structure, which is stacked in the \emph{ABC} order~\cite{zhang2018topological}. The $\mathcal{P}$ is preserved, with the inversion center at the middle Te site of each QL. Once the AFM order is considered, both $\mathcal{T}$ and  $\mathcal{P}$ are broken, but their combination $\mathcal{PT}$ is preserved.

The total energies of different magnetic configurations are calculated, shown in Fig.~S1 of the Suppplemental Material (SM)~\cite{Supp}. The (001)AFM state is found to be the magnetic ground state of Mn$_{2}$Bi$_{6}$Te$_{11}$. The band structures of the (001)AFM state without and with spin-orbit coupling (SOC) are presented in Figs.~3(a) and 3(b), respectively. The fat band structures demonstrate that the highest valence bands and the lowest conduction bands are dominated by $p$ orbitals of Bi and Te atoms. The SOC induces the ``band inversion'' between the valence and conduction bands. The inverted band gap is about 0.18 eV, as seen from the density of states shown in Fig.~3(c). Furthermore, in Figs.~3(d) and 3(e), we plot the evolution of the energy gap with gradually tuning the SOC by $\lambda\ast\lambda_0$, where $\lambda_0$ denotes the realistic SOC strength. The energy gap first decreases to a minimum at about $\lambda=0.68$ [see the inset in Fig.~3(e)] where the ``band inversion'' occurs, and then increases again. The minimum gap $\sim7.6$ meV results from the mass term induced by the AFM order.

The electronic structure of Mn$_{2}$Bi$_{6}$Te$_{11}$ can be captured by the above low-energy effective model, and the detailed parameters are provided in the SM~\cite{Supp}, which can be obtained by fitting with band structures from first-principles calculations. The negative value of $M$ describes the inverted band structure. To investigate the surface states of Mn$_{2}$Bi$_{6}$Te$_{11}$, we employed maximally localized Wannier functions~\cite{Marzari1997, Zhang2009Electronic}, shown in Fig.~S3(a)~\cite{Supp}, where the MnBi$_{2}$Te$_{4}$ SL is chosen as the surface termination. The expected gapped Dirac surface state is clearly seen. We also calculate the Fermi surface in Fig.~S3(b)~\cite{Supp}, which is consistent with the effective model analysis. When the energy level is increased, $k$-cubic terms must be taken into account, leading to a triangular Fermi surface, as shown in Fig.~S3(c)~\cite{Supp}, in contrast to the hexagonal Fermi surface in Bi$_{2}$Te$_{3}$, since the $\mathcal{T}$ is broken.\\

Mn$_{2}$Bi$_{6}$Te$_{11}$ has both inverted bulk bands and gapped surface states, and expects to realize the TI-based dynamical axion state with a hidden pseudospin Chern quantity $\nu=-1$. Moreover, based on the low-energy model, we have explicitly calculated the static part $\theta_0$~\cite{li2010dynamical, Supp} of the axion field as a function of the SOC strength from $\lambda=0$ to $\lambda=1$, as shown by the blue line in Fig. 4(a). Due to the presence of the $m_{5}$ term, $\theta_0$ deviates from $0$ and $\pi$. With increasing the SOC strength, $\theta_0$ continuously changes from near $0$ with $M\gg |m_{5}|$ (NI-based phase) to a large value $\sim\pi$ with $M\ll-|m_{5}|$ (TI-based phase). A rapid increase of $\theta_0$ is found around the (crossover) transition point near $\lambda=0.68$ with a minimum gap, which is consistent with the first-principles calculations in Fig.~3(e). Intriguingly, in the presence of AFM fluctuations, the axion field will become dynamical with $\theta=\theta_{0}+\delta\theta(\mathbf{x},t)$, and to linear order, $\delta\theta(\mathbf{x},t)=\delta m_{5}/g$, where $\delta m_{5}$ is proportional to the amplitude fluctuation of AFM order along the $z$ direction, and $g$ is the coefficient~\cite{li2010dynamical}.\\

\begin{figure}
  \centering
  \includegraphics[width=3.2in]{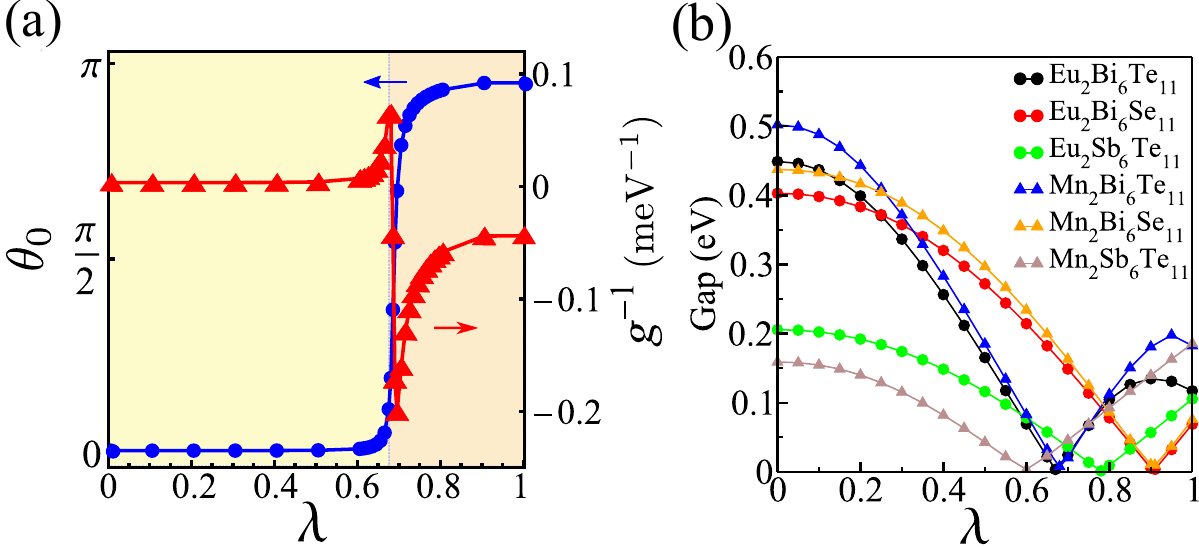}\\
  \caption{(a) The evolution of the static axion angle $\theta_0$ (blue line) and $g^{-1}$ (red line) of Mn$_{2}$Bi$_{6}$Te$_{11}$ with tuning the SOC strength $\lambda$ from 0 to 1, showing a rapid change near the transition point around $\lambda=0.68$. (b) Evolutions of the energy gap with increasing the SOC strength for this family superlattices of $X_{2}Y_{6}Z_{11}$ ($X$=Eu/Mn, $Y$=Bi/Sb, and $Z$=Se/Te). Insets of the energy gaps near the minimum point in (b) are shown in Fig. S5~\cite{Supp}. }\label{fig4}
\end{figure}

\emph {Discussion.} The axion electrodynamics may lead to various exotic physical effects~\cite{Wilczek1987, Sekine2016Chiral, Zhang2019Dynamical,Ooguri2012Insta, imaeda2019axion,li2010dynamical}, which in turn makes it possible to experimentally detect the dynamical axion field. For example, the axionic polariton, as a collective mode emerging from the coupling between the light and axionic mode, can be detected by the attenuated total reflection measurement \cite{li2010dynamical}. To facilitate the experimental detection of the dynamical axion state, a large proportionality coefficient $g^{-1}$ between $\delta\theta$ and the AFM fluctuation $\delta m_{5}$ is highly preferred. In Fig. 4(a), we plot $g^{-1}$ as a function of the SOC strength for Mn$_{2}$Bi$_{6}$Te$_{11}$ (see the red line). It clearly shows that the value of $g^{-1}$ in the TI-based side ($M<0$) is generically one order of magnitude larger than that in the NI-based side ($M>0$). Notably, $g^{-1}$ becomes significantly large when approaching the transition point from the TI-based side ($M\rightarrow 0^{-}$)\cite{Zhang2019Dynamical}. This can be achieved via a chemical replacement by lighter elements with weaker SOC (e.g., Bi by Sb, Te by Se). We present the detailed band structures (see Fig. S4~\cite{Supp}) and gap evolution with tuning the SOC strength in Fig.~4(b) (insets around minimum energy gaps are shown in Fig. S5~\cite{Supp}) for the family superlattices $X_{2}Y_{6}Z_{11}$ ($X$=Eu/Mn, $Y$=Bi/Sb, and $Z$=Se/Te), some of which are also promising candidates to realize the TI-based dynamical axion state.

\emph {Acknowledgement.} H.Z. was supported by the National Natural Science Foundation of China (Grants No.~11674165 and 11834006), the Fok Ying-Tong Education Foundation of China (Grant No.~161006), the Fundamental Research Funds for the Central Universities (Grant No.~021314380147) and High-Performance Computing Center of Collaborative Innovation Center of Advanced Microstructures. J.W. is supported by the Natural Science Foundation of China through Grant No. 11774065, the National Key Research Program of China under Grant No. 2016YFA0300703, and the Natural Science Foundation of Shanghai under Grant Nos. 17ZR1442500, and 19ZR1471400.

H.W. and D.W. contributed equally to this work.

\bibliography{reference}
\end{document}


\title{Supplemental Materials for ``Dynamical axion state with hidden pseudospin Chern numbers in MnBi$_{2}$Te$_{4}$-based heterostructures''}

\author{Huaiqiang Wang$^{1,\dagger}$, Dinghui Wang$^{1,\dagger}$, Zhilong Yang$^{1}$, Minji Shi$^1$, Jiawei Ruan$^1$, Dingyu Xing$^{1,2}$, Jing Wang$^{3,2,4\ast}$, Haijun Zhang$^{1,2\ast}$}

\affiliation{
$^1$ National Laboratory of Solid State Microstructures, School of Physics, Nanjing University, Nanjing 210093, China\\
$^2$ Collaborative Innovation Center of Advanced Microstructures, Nanjing University, Nanjing 210093, China\\
$^3$ State Key Laboratory of Surface Physics, Department of Physics, Fudan University, Shanghai 200433, China\\
$^4$ Institute for Nanoelectronic Devices and Quantum Computing, Fudan University, Shanghai 200433, China
}

\email{zhanghj@nju.edu.cn; wjingphys@fudan.edu.cn}
\maketitle

\tableofcontents

\section{Constraint of $\mathcal{PT}$ symmetry on $\Gamma$ matrices}
A general $4\times 4$ Hamiltonian $\mathcal{H}(\mathbf{k})$ can always be expanded by sixteen linearly independent $4\times 4$ Hermitian matrices as
\begin{equation}
\mathcal{H}(\mathbf{k})=\sum_{i,j=0,x,y,z}d_{ij}(\mathbf{k})\tau_{i}\otimes\sigma_{j},\tag{S1}
\end{equation}
where $\tau_{i}$ and $\sigma_{j}$ are Pauli matrices, and $\tau_{0}$ and $\sigma_{0}$ are $2\times 2$ identity matrices. The $\mathcal{PT}$ symmetry requires that
\begin{equation}
\mathcal{PT}\mathcal{H}(\mathbf{k})(\mathcal{PT})^{-1}=\mathcal{H}(\mathbf{k}),\tag{S2}
\end{equation}
which leads to
\begin{equation}
\mathcal{PT}(\tau_{i}\otimes\sigma_{j})(\mathcal{PT})^{-1}=\tau_{i}\otimes\sigma_{j}.\tag{S3}
\end{equation}
By inserting the explicit form of the $\mathcal{PT}$ operator $\tau_{z}\otimes i\sigma_{y}K$ ($K$ denotes the complex conjugate) of the main text into the above equation, we obtain that only the identity matrix $\mathbb{I}_{4\times 4}$ and the following five $4\times 4$ Hermitian matrices
\begin{equation}
\tau_{x}\otimes\sigma_{x},\tau_{x}\otimes\sigma_{y},\tau_{y}\otimes \sigma_{0}, \tau_{z}\otimes \sigma_{0},\tau_{x}\otimes\sigma_{z}\tag{S4}
\end{equation}
are allowed by $\mathcal{PT}$ symmetry. These five matrices anticommute with each other and exactly correspond to the five Dirac matrices $\Gamma^{i=1,2,3,4,5}$ in the main text.
\section{Derivation of the pseudospin Chern number}
In this section, we present the detailed calculation of the pseudospin Chern number. We begin by rewriting the total effective Hamiltonian $\mathcal{H}(\mathbf{k})$ of the continuum model in the main text in the ordered basis of ($|P1^{+}_{z},\uparrow\rangle$, $|P1^{+}_{z},\downarrow\rangle$, $|P2^{-}_{z},\uparrow\rangle$, $|P2^{-}_{z},\downarrow\rangle$) as
\begin{equation}
\label{continuumH}
\begin{split}
\mathcal{H}(\mathbf{k})=&\tau_{x}(A_{2}k_{y}\sigma_{x}-A_{2}k_{x}\sigma_{y}+m_{5}\sigma_{z})\\
&+A_{1}k_{z}\tau_{y}+M(\mathbf{k})\tau_{z},
\end{split}\tag{S5}
\end{equation}
where the constant energy term has been dropped, $M(\mathbf{k})=M+B_{1}k_{z}^{2}+B_{2}k_{\parallel}^{2}$ with $k_{\parallel}=\sqrt{k_{x}^{2}+k_{y}^{2}}$, and the Pauli matrices $\tau_{i}$ and $\sigma_{i}$ with $i=x,y,z$, act in the orbital and spin subspaces, respectively. For simplicity and without loss of generality, $A_{1}$, $A_{2}$, $B_{1}$, $B_{2}$, and $m_{5}$, are assumed to be positive. The energy spectrum is then easily obtained as
\begin{equation}
\label{spectrum}
E_{\mathrm{c}/\mathrm{v}}(\mathbf{k})=\pm\sqrt{A_{2}^{2}k_{\parallel}^{2}+A_{1}^2k_{z}^{2}+M(\mathbf{k})^{2}+m_{5}^{2}},\tag{S6}
\end{equation}
where c (v) stands for the doubly-degenerate conduction (valence) bands. Note that the band gap remains open regardless of the value of $m_{5}$. For simplicity, we will first calculate the pseudospin Chern number in the absence of $m_{5}$ term which applies to the three-dimensional (3D) topological insulator, and then we will show that the results remain unchanged in the presence of $m_{5}$ term, since it cannot close the pseudospin spectrum gap, which is required to change the pseudospin Chern number.

The definition of pseudospin Chern number stems from a smooth decomposition of the occupied valence bands into two sectors through projecting a properly chosen pseudospin operator onto them as $P_{v}\hat{S}P_{v}$ with $P_{v}$ being the projection operator. Without $m_{5}$, the wavefunctions of the valence bands of Eq. (\ref{continuumH}) can be written as a direct product \cite{Li2010Chern, Yang2011}

\begin{equation}
\label{wave}
\psi_{\pm}(\mathbf{k})=\chi_{\pm}(\mathbf{k})\otimes\phi_{\pm}(\mathbf{k}),\tag{S7}
\end{equation}
where
\begin{equation}
\label{wavefunction}
\begin{split}
\phi_{\pm}(\mathbf{k})=&\frac{1}{\sqrt{2}}\big[1,\ \mp i e^{i\theta_{k}}\big]^{\mathrm{T}}, \\
\chi_{\pm}(\mathbf{k})=&\Big[\sin\frac{\alpha_{k}}{2},\mp e^{\pm i\beta_{k}}\cos\frac{\alpha_{k}}{2}\Big]^{\mathrm{T}},
\end{split}\tag{S8}
\end{equation}
are two-component wavefunctions in the spin and orbital subspaces, respectively. Here,
\begin{equation}
\begin{split}
e^{i\theta_{k}}=&\frac{k_{x}+ik_{y}}{k_{\parallel}},\\
\alpha_{k}=&\arccos\frac{M(\mathbf{k})}{E(\mathbf{k})},\\
e^{i\beta_{k}}=&\frac{A_{2}k_{\parallel}+iA_{1}k_{z}}{\sqrt{ A_{2}^{2}k_{\parallel}^{2} +A_{1}^{2}k_{z}^{2}}},
\end{split}\tag{S9}
\end{equation}
where $E(\mathbf{k})=\sqrt{ A_{2}^{2}k_{\parallel}^{2} +A_{1}^{2}k_{z}^{2}+M(\mathbf{k})^{2}}$.

Based on the fact that the pseudospin Chern number can only be defined in a two-dimensional (2D) system, we can treat one of the three momenta $k_{x}, k_{y}$, and $k_{z}$ as a parameter to reduce the noninteracting 3D system into a collection of decoupled 2D slices.

\subsection{$k_{z}$-dependent pseudospin Chern number}

First we consider $k_{z}$ as a parameter, and the pseudospin operator can be constructed by inspecting the time-reversal-invariant $k_{z}=0$ plane as follows. It should be emphasized that TRS is not a must here, since the pseudospin Chern number remains intact regardless of TRS \cite{Yang2011}. The intention of including TRS here is to simplify the process of looking for a proper pseudospin operator.  When $k_{z}=0$, via a basis transformation through the unitary matrix
\begin{equation}
U_{z}=\left(
        \begin{array}{cccc}
          1 & 0 & 0 & 0 \\
          0 & 0 & 0 & 1 \\
          0 & 1 & 0 & 0 \\
          0 & 0 & -1 & 0 \\
        \end{array}
      \right),\tag{S10}
\end{equation}
the Hamiltonian can be block-diagonalized as
\begin{equation}
\label{BHZ}
U_{z}\mathcal{H}(k_{z}=0)U_{z}^{\dagger}=\left(
                        \begin{array}{cc}
                          h_{z}(\mathrm{\mathbf{k}}) & 0 \\
                          0 & h_{z}(-\mathrm{\mathbf{k}})^{*} \\
                        \end{array}
                      \right),\tag{S11}
\end{equation}
where
\begin{equation}
h_{z}(\mathbf{\mathbf{k}})=(M+B_{2}k_{\parallel}^{2})\sigma_{z}+A_{2}k_{y}\sigma_{x}-A_{2}k_{x}\sigma_{y},\tag{S12}
\end{equation}
and $h_{z}(-\mathbf{k})^{*}$ is connected to $h_{z}(\mathbf{k})$ by TRS. The Hamiltonian in Eq. (\ref{BHZ}) resembles the celebrated Bernevig-Hughes-Zhang (BHZ) model \cite{bernevig2006quantum} for quantum spin Hall effect with opposite Chern numbers for $h_{z}(\mathbf{k})$ and $h_{z}(-\mathbf{k})^{*}$, respectively, in the inverted-band regime ($M/B_{2}<0$). Consequently, we can naturally define a pseudospin operator in the above transformed basis as $\tau_{z}$ (If $-\tau_{z}$ is used here, then the result of the pseudospin Chern number will be reversed). In the original basis, it is represented as
\begin{equation}
\hat{S}=U_{z}^{\dagger}\tau_{z}U_{z}=\tau_{z}\sigma_{z}.\tag{S13}
\end{equation}

By projecting $\hat{S}$ to the valence bands subspace on the $|\psi_{+}\rangle$, $|\psi_{-}\rangle$ basis, we get
\begin{equation}
P_{\mathrm{v}}\hat{S}P_{\mathrm{v}}=\left(
                                      \begin{array}{cc}
                                        \langle\psi_{+}|\hat{S}|\psi_{+}\rangle & \langle\psi_{+}|\hat{S}|\psi_{-}\rangle  \\
                                        \langle\psi_{-}|\hat{S}|\psi_{+}\rangle  & \langle\psi_{-}|\hat{S}|\psi_{-}\rangle  \\
                                      \end{array}
                                    \right)=\left(
                                              \begin{array}{cc}
                                                0 & t_{k} \\
                                                t_{k}^{*} & 0 \\
                                              \end{array}
                                            \right),\tag{S14}
\end{equation}
where
\begin{equation}
t_{k}=e^{-2i\beta_{k}}\cos^{2}\frac{\alpha_{k}}{2}+\sin^{2}\frac{\alpha_{k}}{2}.\tag{S15}
\end{equation}
The pseudospin spectrum can then be obtained as
\begin{equation}
\epsilon_{S\pm}(\mathbf{k})=\pm |t_{k}|,\tag{S16}
\end{equation}
with corresponding eigenstates
\begin{equation}
\varphi^{S}_{\pm}(\mathbf{k})=\frac{1}{\sqrt{2}}\Big[\psi_{+}(\mathbf{k})\pm \frac{t_{k}^{*}}{|t_{k}|} \psi_{-}(\mathbf{k})\Big].\tag{S17}
\end{equation}
When the pseudospin is conserved (e.g., when $k_{z}=0$), the pseudospin spectrum always equals $\pm1$, as it should be. However, when the pseudospin conservation is broken (e.g., when $k_{z}\neq0$), the pseudospin spectrum will generally deviate from $\pm1$, leading to the formation of two islands around $+1$ and $-1$, respectively, with a gap separating them. As long as the pseudospin gap exists in the whole $k_{x}$-$k_{y}$ slice, a usual Chern number can be defined for each branch~\cite{Li2010Chern}
\begin{equation}
\label{spin Chern}
C_{\pm}^{S}(k_{z})=\frac{1}{2\pi}\int d^{2}k\Omega_{\pm}^{xy}(k_{z}),\tag{S18}
\end{equation}
where $\Omega_{\pm}^{xy}(k_{z})=i\mathbf{\hat{e}_{z}}\cdot[\nabla_{k}\times\langle\varphi_{\pm}^{S}|\nabla_{k}|\varphi_{\pm}^{S}\rangle]$ is the Berry curvature in the $k_{x}$-$k_{y}$ plane at a fixed $k_{z}$ value. The pseudospin Chern number is defined as $C^{S}=(C_{+}^{S}-C_{-}^{S})/2$.

Before the detailed calculation of $C^{S}(k_{z})$, we look for the gap-closing conditions for the pseudospin spectrum, since the change of $C^{S}$ requires a gap closing-and-reopening process. The pseudospin spectrum gap closes when $t_{k}=0$, leading to
\begin{equation}
(\alpha_{k},\beta_{k})=\Big(\frac{\pi}{2},\frac{\pi}{2}\Big)\ \mathrm{or} \ \Big(\frac{\pi}{2},\frac{3\pi}{2}\Big),\tag{S19}
\end{equation}
which are satisfied at $(k_{x},k_{y},k_{z})=(0,0,\pm k_{zc})$, with $k_{zc}=\sqrt{-M/B_{1}}$. By expressing the Berry curvature in the polar coordinate in the $k_{x}$-$k_{y}$ plane, the Berry curvature can be simplified as~\cite{Li2010Chern}
\begin{equation}
\Omega_{\pm}(k_{z})=\frac{1}{2k_\parallel}\frac{\partial}{\partial k_\parallel}P_{\pm}(k_{\parallel},k_{z}),\tag{S20}
\end{equation}
where $P_{\pm}(k_{\parallel},k_{z})$ is given by
\begin{equation}
P_{\pm}(k_\parallel,k_{z})=2i\langle\varphi_{\pm}^{S}|\frac{\partial}{\partial\theta_{k}}|\varphi_{\pm}^{S}\rangle.\tag{S21}
\end{equation}
The $k_{z}$-dependent Chern number for each pseudospin sector can then be derived as
\begin{equation}
\label{Csz}
\begin{split}
C^{S}_{\pm}(k_{z})=&\frac{1}{2}[P_{\pm}(\infty,k_{z})-P_{\pm}(0,k_{z})]\\
=&\frac{1}{2}\Big[\big(-1\mp \mathrm{sgn}(B_{2})\big)-\big(-1\mp \mathrm{sgn}(M+B_{1}k_{z}^{2})\big)\Big]\\
=&\pm\frac{1}{2}\big[-\mathrm{sgn}(B_{2})+\mathrm{sgn}(M+B_{1}k_{z}^{2})\big].
\end{split}\tag{S22}
\end{equation}
Since $B_{1}$ and $B_{2}$ are positive, and $M$ is negative for the inverted band structure in our model, the above expression leads to
\begin{equation}
C^{S}(k_{z})=\frac{C_{+}^{S}-C_{-}^{S}}{2}=\left\{
                     \begin{array}{ll}
                       -1, & \hbox{$|k_{z}|<\sqrt{-M/B_{1}}$;} \\
                       0, & \hbox{$|k_{z}|>\sqrt{-M/B_{1}}$.}
                     \end{array}
                   \right.\tag{S23}
\end{equation}
Note that if $M>0$ (band structure in the normal order), $C^{S}(k_{z})$ always equals zero. Furthermore, as expected, the boundaries between distinct regions of $C^{S}(k_{z})$ are consistent with the gap closing conditions of the pseudospin spectrum at $k_{z}=\pm\sqrt{-M/B_{1}}$. In addition, the total Chern number of the valence bands is given by
\begin{equation}
C(k_{z})=C_{+}^{S}(k_{z})+C_{-}^{S}(k_{z})=0.\tag{S24}
\end{equation}

Now we consider the effect of the AFM $m_{5}$ term on the pseudospin spectrum. With this term, the wavefunction in Eq. (\ref{wavefunction}) becomes
\begin{equation}
\label{wavefunction2}
\begin{split}
\phi_{+}(\mathbf{k})=&\Big[\cos\frac{\eta_{k}}{2}, -ie^{i\theta_{k}}\sin\frac{\eta_{k}}{2}\Big]^{\mathrm{T}}, \\
\phi_{-}(\mathbf{k})=&\Big[\sin\frac{\eta_{k}}{2}, ie^{i\theta_{k}}\cos\frac{\eta_{k}}{2}\Big]^{\mathrm{T}}, \\
\chi_{\pm}(\mathbf{k})=&\Big[\sin\frac{\alpha_{k}'}{2},\mp e^{\pm i\beta_{k}'}\cos\frac{\alpha_{k}'}{2}\Big]^{\mathrm{T}},
\end{split}\tag{S25}
\end{equation}
with
\begin{equation}
\begin{split}
\eta_{k}=&\arccos\frac{m_{5}}{\sqrt{A_{2}^{2}k_{\parallel}^{2}+m_{5}^{2}}},\\
\alpha_{k}'=&\arccos\frac{M(\mathbf{k})}{\sqrt{A_{2}^{2}k_{\parallel}^{2}+A_{1}^2k_{z}^{2}+M(\mathbf{k})^{2}+m_{5}^{2}}},\\
e^{i\beta_{k}'}=&\frac{\sqrt{A_{2}^{2}k_{\parallel}^{2}+m_{5}^{2}}+iA_{1}k_{z}}{\sqrt{A_{2}^{2}k_{\parallel}^{2}+A_{1}^2k_{z}^{2}+M(\mathbf{k})^{2}+m_{5}^{2}}}.
\end{split}\tag{S26}
\end{equation}
After some algebra, the pseudospin operator can be derived as
\begin{equation}
P_{\mathrm{v}}\hat{S}P_{\mathrm{v}}=\left(
                                             \begin{array}{cc}
                                                -\cos(\eta_{k})\cos(\alpha_{k}') & e^{-i\beta_{k}'}\sin(\eta_{k})b_{k} \\
                                                e^{i\beta_{k}'}\sin(\eta_{k})b_{k}^{*} &\cos(\eta_{k})\cos(\alpha_{k}') \\
                                              \end{array}
                                            \right),\tag{S27}
\end{equation}
where $b_{k}=\cos\beta_{k}'-i\cos(\alpha_{k}')\sin(\beta_{k}')$. It is found that in the presence of $m_{5}$ the pseudospin spectrum get modified, but the gap is still closed in the $k_{z}=\pm \sqrt{-M/B_{1}}$ slices at $(k_{x},k_{y})=(0,0)$ and remains open for other $k_{z}$ slices. Consequently, according to the adiabatic continuity argument, the pseudospin Chern number $C^{S}_\pm(k_{z})$ keeps the same as that in the absence of $m_{5}$ term obtained above. Nevertheless, we have also carried out the detailed calculation as in Eq. (\ref{Csz}), which leads to the same expression $C^{S}(k_{z})=(1/2)\big[-\mathrm{sgn}(B_{2})+\mathrm{sgn}(M+B_{1}k_{z}^{2})\big]$ as expected.

\subsection{$k_{x (\mathrm{or}\ y)}$-dependent pseudospin Chern number}
Then we investigate the $k_{x}$- and $k_{y}$- dependent pseudospin Chern numbers $C^{S}(k_{x})$ and $C^{S}(k_{y})$ by treating $k_{x}$ and $k_{y}$ as parameters, respectively.

Similar to the above analysis, by inspecting the $k_{x}=0$ plane, the transformation matrix can be chosen as
\begin{equation}
U_{x}=\frac{1}{\sqrt{2}}\left(
        \begin{array}{cccc}
          1 & 1 & 0 & 0 \\
          0 & 0 & 1 & 1 \\
          1 & -1 & 0 & 0 \\
          0 & 0 & 1 & -1 \\
        \end{array}
      \right),\tag{S28}
\end{equation}
which block-diagonalizes $\mathcal{H}(k_{x}=0)$ to a BHZ-type Hamiltonian in the absence of $m_{5}$ term as
\begin{equation}
U_{x}\mathcal{H}(k_{x}=0)U_{x}^{\dagger}=\left(
                        \begin{array}{cc}
                          h_{x}(\mathrm{\mathbf{k}}) & 0 \\
                          0 & h_{x}(-\mathrm{\mathbf{k}})^{*} \\
                        \end{array}
                      \right),\tag{S29}
\end{equation}
with
\begin{equation}
h_{x}(\mathbf{\mathbf{k}})=(M+B_{1}k_{z}^{2}+B_{2}k_{y}^{2})\sigma_{z}+A_{2}k_{y}\sigma_{x}+A_{1}k_{z}\sigma_{y},\tag{S30}
\end{equation}
which has a Chern number of $+1$ in the inverted regime of $M<0$, while $h_{x}(-\mathbf{k})^{*}$ has an opposite Chern number. The pseuospin operator $\tau_{z}$ is then transformed back to
\begin{equation}
\hat{S}=U_{x}^{\dagger}(\tau_{z}\sigma_{0})U_{x}=\tau_{0}\sigma_{x}.\tag{S31}
\end{equation}
By taking an analogous procedure as in the $k_{z}$ case (the calculation becomes simpler in the transformed basis), it is straight forward to obtain the $k_{x}$-dependent Chern number for each pseudospin sector in $k_{y}$-$k_{z}$ slices both without and with $m_{5}$ term as
\begin{equation}
\label{Csx}
C^{S}_{\pm}(k_{x})=\pm\frac{1}{2}\big[-\mathrm{sgn}(B_{1})+\mathrm{sgn}(M+B_{2}k_{x}^{2})\big].\tag{S32}
\end{equation}
where $B_{2}$ and $B_{1}$ are assumed positive. When $M<0$ (inverted band structure), this leads to
\begin{equation}
\label{Csx2}
C^{S}(k_{x})=\left\{
                     \begin{array}{ll}
                       -1, & \hbox{$|k_{x}|<\sqrt{-M/B_{2}}$;} \\
                       0, & \hbox{$|k_{x}|>\sqrt{-M/B_{2}}$.}
                     \end{array}
                   \right.\tag{S33}
\end{equation}
When $M>0$, all $k_{y}$-$k_{z}$ slices are topologically trivial with $C^{S}(k_{x})=0$

Similarly, when treating $k_{y}$ as a parameter, the transformation matrix and pseudospin operators are given by
\begin{equation}
U_{y}=\frac{1}{\sqrt{2}}\left(
        \begin{array}{cccc}
          i & 1 & 0 & 0 \\
          0 & 0 & i & 1 \\
          1 & i & 0 & 0 \\
          0 & 0 & 1 & i \\
        \end{array}
      \right),\quad\hat{S}=U_{y}^{\dagger}(\tau_{z}\sigma_{0})U_{y}=\tau_{0}\sigma_{y}.\tag{S34}
\end{equation}
After some algebra, the results for $C^{S}(k_{y})$ turns out to be the same as Eqs. (\ref{Csx}) and (\ref{Csx2}) upon replacing $k_{x}$ by $k_{y}$.

\section{Wilson-loop characterization}
In this section, we adopt the Wilson-loop method \cite{Yu2011Equivalent} to supplement the quantized property associated with the above pseudospin Chern number. In contrast to the conventional Wilson loops defined from the ground state of real band structures, here we focus on the bands in the pseudospin spectrum and treat the lower one of the two pseudospin branches as the ``occupied'' valence band. The Wilson-loop operator for the single ``occupied'' band is then simply given by
\begin{equation}
\label{Wilsonloop}
\mathcal{W}(\mathcal{L})=\langle u(\mathbf{k}^{0}+\mathbf{G})|\overline{\prod}_{\{\mathbf{k}^{i}\}_{i=1}^{N}\in \mathcal{L}} \mathcal{P}^{\mathrm{occ}}(\mathbf{k}^{i})|u(\mathbf{k}^{0})\rangle,\tag{S35}
\end{equation}
where $|u(\mathbf{k})\rangle$ is the eigenstate of the ``occupied'' band, and $\mathcal{P}^{\mathrm{occ}}(\mathbf{k})=|u(\mathbf{k})\rangle\langle u(\mathbf{k})|$ is the projection operator to the ``occupied'' band, and the bar in $\overline{\prod}$ means path ordering. The unitary Wilson-loop operator has a single eigenvalue $e^{i\gamma}$, where $\gamma$ corresponds to the $U(1)$ Berry phase.

As a concrete illustration, we first treat $k_{z}$ as a parameter to investigate the evolution of $\gamma(k_{x})$ in the 2D $k_{x}$-$k_{y}$ slice by choosing the Wilson-loop along the $k_{y}$ direction. Note that to calculate the Wilson-loop operator,
we must transform the above continuum model to a lattice one, and the Chern number of each pseudospin sector $C_{\pm}^{S}(k_{z})$ changes to
\begin{equation}
C_{\pm}^{S}(k_{z})=\left\{
                     \begin{array}{ll}
                       \mp1, & \hbox{$\cos k_{z}>1+\frac{M}{2B_{1}}$;} \\
                       0, & \hbox{$\cos k_{z}<1+\frac{M}{2B_{1}}$.}
                     \end{array}
                   \right.\tag{S36}
\end{equation}
where $(-4B_{1}<M<0)$ is assumed, and $C_{\pm}^{S}(k_{z})=0$ for other values of $M$.

As shown by the blue line in Fig. 1(b), when $\cos k_{z}>1+M/2B_{1}$, the Wilson-loop spectra $\gamma(k_{x})$ exhibits a nontrivial winding along $k_{x}$, which corresponds to a nonzero pseudospin Chern number $C_{-}^{S}(k_{z})=+1$, while when $\cos k_{z}<1+M/2B_{1}$, there is no winding of $\gamma(k_{x})$ [see the red line in Fig. 1(b)], corresponding to $C_{-}^{S}(k_{z})=0$. When treating $k_{x}$ or $k_{y}$ as a parameter, the Wilson-loop spectra in the $k_{y}$-$k_{z}$ or $k_{x}$-$k_{z}$ slice exhibits similar behaviors as above, which will thus not be shown.

\section{Effective model Hamiltonians for surface states}
In this section, we explicitly derive the low-energy effective Hamiltonians of the surface states in the presence of the AFM order. We start from the lattice version of the continuum model in the main text, which can be described by the Hamiltonian $H=\sum_{\mathbf{k}}\mathcal{H}(\mathbf{k})c_{\mathbf{k}}^{\dagger}c_{\mathbf{k}}$ with
\begin{equation}
\label{dkgamma}
\begin{split}
\mathcal{H}(\mathbf{k})=&\epsilon_{0}(\mathbf{k})+\sum^{5}_{i=1}d_{i}(k)\Gamma^{i},\\
d_{1,2,...,5}(\mathbf{k})=&\Big(A_{2}\sin k_{y},-A_{2}\sin k_{x},A_{1}\sin k_{z},M(\mathbf{k}),m_{5} \Big),
\end{split}\tag{S37}
\end{equation}
where $\epsilon_{0}(\mathbf{k})=C+2D_{1}+4D_{2}-2D_{1}\cos k_{z}-2D_{2}(\cos k_{x}+\cos k_{y})$, $M(\mathbf{k})=M+2B_{1}+4B_{2}-2B_{1}\cos k_{z}-2B_{2}(\cos k_{x}+\cos k_{y})$, and the five Dirac matrices are represented as $\Gamma^{1,2,...,5}=(\tau_{x}\otimes\sigma_{x},\tau_{x}\otimes\sigma_{y},\tau_{y}\otimes \sigma_{0}, \tau_{z}\otimes \sigma_{0},\tau_{x}\otimes\sigma_{z})$ in the ordered basis of ($|P1^{+}_{z},\uparrow\rangle$, $|P1^{+}_{z},\downarrow\rangle$, $|P2^{-}_{z},\uparrow\rangle$, $|P2^{-}_{z},\downarrow\rangle$). For later convenience, we also define $\Gamma^{ij}=(1/2i)[\Gamma^{i},\Gamma^{j}]$. For simplicity, we will neglect the constant energy term $\epsilon_{0}(\mathbf{k})$ in the following discussions.

\subsection{surface states in $z$-direction}
First we consider the surface states along the $z$ direction by taking periodic boundary condition (PBC) in $x$- and $y$- directions and open boundary condition (OBC) in $z$-direction. By treating the good quantum numbers $k_{x}$ and $k_{y}$ as parameters and Fourier transforming $k_{z}$ into position space through the substitution $c_{\mathbf{k}}=(1/L)\sum_{z}e^{ik_{z}z}c_{k_{x},k_{y},z}$, we can obtain an effective reduced one-dimensional tight-binding Hamiltonian as
\begin{equation}
\label{realH}
\begin{split}
H(k_{x},k_{y})=&\sum_{z}\Big(\mathcal{M}c_{k_{x},k_{y},z}^{\dagger}c_{k_{x},k_{y},z}+\mathcal{\mathcal{N}}c_{k_{x},k_{y},z}^{\dagger}c_{k_{x},k_{y},z+1}\\
&+\mathcal{\mathcal{N}}^{\dagger}c_{k_{x},k_{y},z+1}^{\dagger}c_{k_{x},k_{y},z}\Big),\\
\mathcal{M}=&\Big[M+2B_{1}+4B_{2}-2B_{2}(\cos k_{x}+\cos k_{y})\Big]\Gamma^{4}\\
&+A_{2}(\sin k_{y}\Gamma^{1}-\sin k_{x}\Gamma^{2}),\\
\mathcal{N}=&\frac{i}{2}A_{1}\Gamma^{3}-B_{1}\Gamma^{4}.
\end{split}\tag{S38}
\end{equation}
Considering the exponentially localized nature of the surface states, we can take the ansatz $\psi_{\alpha}(z)=\lambda^{z}\phi_{\alpha}$, where $\phi$ is a four-component spinor with $\alpha=1,...,4$, $\lambda$ is a complex number, and $z$ denotes the lattice site from $z=0$ to $L$. Then the real-space eigenequation $H\psi=E\psi$ amounts to
\begin{equation}
\label{eigeneq}
\lambda \mathcal { N } _ { \alpha \beta } \phi _ { \beta } + \lambda ^{-1} \mathcal { N } _ { \alpha \beta } ^ { \dagger } \phi _ { \beta } + \mathcal { M } _ { \alpha \beta } \phi _ { \beta } = E \phi _ { \alpha }.\tag{S39}
\end{equation}
In the absence of the AFM $m_{5}\Gamma_{5}$ term, the system is both time-reversal (TR) and particle-hole symmetric, and consequently a zero-energy solution $E=0$ for the surface state is expected at the $\Gamma$ point, leading to the simplified equation
\begin{equation}
\label{simpleeq}
\bigg[\frac{iA_{1}}{2}(\lambda-\lambda^{-1})\Gamma^{3}+\Big(-B_{1}(\lambda+\lambda^{-1})+\mathcal{M}(0)\Big)\Gamma^{4}\bigg]\phi=0,\tag{S40}
\end{equation}
where $\mathcal{M}(0)=M+2B_{1}$. Multiplying both sides by $\Gamma^{4}$ yields
\begin{equation}
\frac{A_{1}}{2}(\lambda-\lambda^{-1})\Gamma^{34}\phi=\Big[-B_{1}(\lambda+\lambda^{-1})+\mathcal{M}(0)\Big]\phi,\tag{S41}
\end{equation}
where $\Gamma^{34}\equiv (1/2i)[\Gamma^{3},\Gamma^{4}]=\tau_{x}\sigma_{0}$, with eigenvalues $\pm1$. For $\Gamma^{34}\phi=\phi$, the eigenvectors are given by
$\phi_{1+}=(1/\sqrt{2})[1,0,1,0]^{\mathrm{T}}$ and $\phi_{2+}=(1/\sqrt{2})[0,1,0,1]^{\mathrm{T}}$, respectively, and $\lambda$ can be solved as
\begin{equation}
\label{lambda}
\lambda_{1,2}=\frac{\mathcal{M}(0)\pm\sqrt{\mathcal{M}(0)^{2}+(A_{1}^{2}-4B_{1}^{2})}}{A_{1}+2B_{1}}.\tag{S42}
\end{equation}
Note that if $\lambda$ is a solution for $\Gamma^{34}\phi=\phi$, then $\lambda^{-1}$ will be the solution for the opposite eigenvalue $\Gamma^{34}\phi=-\phi$, with eigenvectors $\phi_{1-}=(1/\sqrt{2})[-1,0,1,0]^{\mathrm{T}}$ and $\phi_{2-}=(1/\sqrt{2})[0,-1,0,1]^{\mathrm{T}}$. The surface state localized at the bottom (top) surface with $z=0$ ($z=L$) requires $|\lambda_{1,2}|<1$ ($|\lambda_{1,2}|>1$) for a normalizable solution, leading to the condition $-4B_{1}<M<0$, which is consistent with the band inversion condition at $\Gamma$ point in the topological regime. Since $A_{1}/B_{1}>0$ in our model, $|(A_{1}-2B_{1})/(A_{1}+2B_{1})|<1$, and thus we get $|\lambda_{1,2}|<1$ in Eq. (\ref{lambda}), indicating that $\phi_{s+}$ ($\phi_{s-}$) with $s=1,2$, corresponds to the bottom (top) surface state.

As for the bottom surface state, the boundary condition of $\psi_{\alpha}(0)=0$ gives rise to the solution at the $\Gamma$ point
\begin{equation}
\psi_{s+}(z)=(\lambda_{1}^{z}-\lambda_{2}^{z})\phi_{s+}, \quad s=1,2.\tag{S43}
\end{equation}
Away from the $\Gamma$ point and in the presence of the TR-breaking mass term $m_{5}\Gamma^{5}$, the effective Hamiltonian of the bottom surface state can be obtained by expressing these terms in the above basis as a $2\times 2$ matrix as
\begin{equation}
\label{bottomsurface}
h^{z=0}_{\mathrm{surf}}=A_{2}(k_{y}\sigma_{x}-k_{x}\sigma_{y})+m_{5}\sigma_{z}.\tag{S44}
\end{equation}

Similarly, for the top surface state, the effective Hamiltonian can be expanded in the two-dimensional $\phi_{1-}$ and $\phi_{2-}$ subspace as
\begin{equation}
\label{topsurface}
h^{z=L}_{\mathrm{surf}}=-\big[A_{2}(k_{y}\sigma_{x}-k_{x}\sigma_{y})+m_{5}\sigma_{z}\big].\tag{S45}
\end{equation}
Obviously, both the top and bottom surface states describe massive Dirac fermions with the energy dispersion
\begin{equation}
E=\pm\sqrt{A_{2}^{2}(k_{x}^{2}+k_{y}^{2})+m_{5}^{2}},\tag{S46}
\end{equation}
while they exhibit opposite fractional Hall conductance as $\sigma_{b,t}=\pm(e^{2}/2h)\mathrm{sgn}(m_{5})$~\cite{Bernevig2013Topological}, which cancel each other as a manifestation of the axion insulator state.

Furthermore, by projecting the real spin operator into to the surface states, it is found that the $\sigma$ matrix in the effective Hamiltonians of Eqs. (\ref{bottomsurface}) and (\ref{topsurface}) is proportional to the real spin. Consequently, both surface states in the $z$-direction will exhibit helical inplane spin textures, as shown in Figs. S3(b)-(c).

\subsection{surface states in $x$- and $y$- directions}
For the surface state along $x$-direction, similar procedures can be taken, leading to the condition of $\lambda$
\begin{equation}
\frac{A_{2}}{2}(\lambda-\lambda^{-1})\Gamma^{24}\phi=\Big[B_{2}(\lambda+\lambda^{-1})-\mathcal{M}(0)\Big]\phi,\tag{S47}
\end{equation}
where $\Gamma^{24}=-\tau_{y}\sigma_{y}$, and the eigenequation of $\Gamma^{24}\phi_{s\pm}^{x}=\pm\phi_{s\pm}^{x}$ with $s=1,2$, generates four eigenstates
\begin{equation}
\begin{split}
\phi_{1+}^{x}=&\frac{1}{\sqrt{2}}[1,0,0,1]^{\mathrm{T}},\ \phi_{2+}^{x}=\frac{1}{\sqrt{2}}[0,-1,1,0]^{\mathrm{T}}\\
\phi_{1-}^{x}=&\frac{1}{\sqrt{2}}[-1,0,0,1]^{\mathrm{T}},\ \phi_{2-}^{x}=\frac{1}{\sqrt{2}}[0,1,1,0]^{\mathrm{T}}.
\end{split}\tag{S48}
\end{equation}
Similar analysis shows that $\phi_{s+}^{x}$ and $\phi_{s-}^{x}$ states are localized on the $x=L$ and $x=0$ surfaces, respectively. The effective Hamiltonians are then derived as
\begin{equation}
\label{hx}
\begin{split}
h_{\mathrm{surf}}^{x=L}=&A_{2}k_{y}\sigma_{z}+A_{1}k_{z}\sigma_{y}+m_{5}\sigma_{x}\\
h_{\mathrm{surf}}^{x=0}=&-A_{2}k_{y}\sigma_{z}-A_{1}k_{z}\sigma_{y}-m_{5}\sigma_{x},
\end{split}\tag{S49}
\end{equation}
with the massive Dirac spectrum given by
\begin{equation}
E=\pm\sqrt{A_{2}^{2}k_{y}^{2}+A_{1}^{2}k_{z}^{2}+m_{5}^{2}}.\tag{S50}
\end{equation}

Finally, along $y$- direction, we get the condition
\begin{equation}
\frac{A_{2}}{2}(\lambda-\lambda^{-1})\Gamma^{14}\phi=\Big[-B_{2}(\lambda+\lambda^{-1})+\mathcal{M}(0)\Big]\phi,\tag{S51}
\end{equation}
with $\Gamma^{14}=-\tau_{y}\sigma_{x}$ and the eigenstates
\begin{equation}
\begin{split}
\phi_{1+}^{y}=&\frac{1}{\sqrt{2}}[i,0,0,1]^{\mathrm{T}},\ \phi_{2+}^{y}=\frac{1}{\sqrt{2}}[0,i,1,0]^{\mathrm{T}}\\
\phi_{1-}^{y}=&\frac{1}{\sqrt{2}}[-i,0,0,1]^{\mathrm{T}},\ \phi_{2-}^{y}=\frac{1}{\sqrt{2}}[0,-i,1,0]^{\mathrm{T}}.
\end{split}\tag{S52}
\end{equation}
The $\phi_{s+}^{y}$ and $\phi_{s-}^{y}$ states are found to be localized on the $y=0$ and $y=L$ surfaces, respectively, and the effective Hamiltonians are given by
\begin{equation}
\label{hy}
\begin{split}
h_{\mathrm{surf}}^{y=L}=&-A_{2}k_{x}\sigma_{z}+A_{1}k_{z}\sigma_{x}-m_{5}\sigma_{y}\\
h_{\mathrm{surf}}^{y=0}=&A_{2}k_{x}\sigma_{z}-A_{1}k_{z}\sigma_{x}+m_{5}\sigma_{y},
\end{split}\tag{S53}
\end{equation}
with the spectrum
\begin{equation}
E=\pm\sqrt{A_{2}^{2}k_{x}^{2}+A_{1}^{2}k_{z}^{2}+m_{5}^{2}}.\tag{S54}
\end{equation}
It is worth mentioning that the AFM $m_{5}\Gamma^{5}$ term gaps all surface Dirac cone states with a gap of 2$m_{5}$, regardless of the surface orientation.

\section{Details of $\mathrm{Mn_{2}Bi_{6}Te_{11}}$}
\subsection{Total energy calculation}
In Fig. S1, we present the total energy obtained from first-principles calculations for different magnetic ordered states of $\mathrm{Mn}_{2}\mathrm{Bi}_{6}\mathrm{Te}_{11}$, namely, the (001) AFM, (100) AFM, (001) FM, and (100) FM, among which the ($001$) AFM is the magnetic ground state.
\begin{figure}
  \centering
  \includegraphics[width=1.8in]{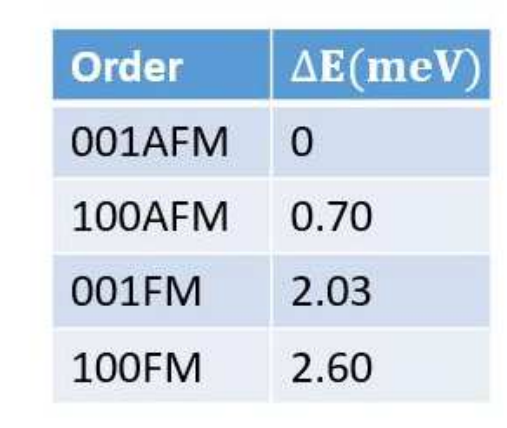}\\
  \caption{Total energies for different magnetic ordered states of $\mathrm{Mn}_{2}\mathrm{Bi}_{6}\mathrm{Te}_{11}$.}\label{figs1}
\end{figure}

\subsection{$\mathbf{k\cdot p}$ parameters}
\begin{figure*}
  \centering
  \includegraphics[width=5.8in]{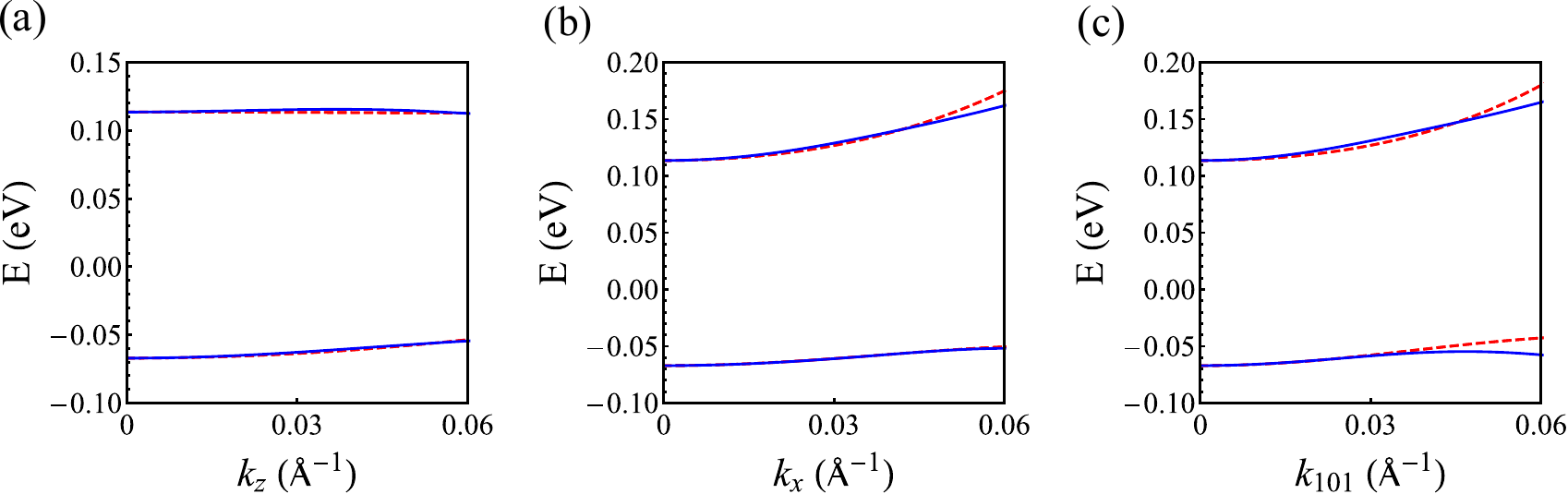}\\
  \caption{Energy dispersions obtained from $k\cdot p$ model (red dashed lines) and \emph{ab initio} calculations (blue solid lines) along (a) $k_{z}$ direction, (b) $k_{x}$ direction, and, (c) $k_{101}$ direction.}\label{figs2}
\end{figure*}

By fitting with band structures from first-principles calculations, the detailed $k\cdot p$ parameters are given by $C=0.0232$ eV, $D_{1}=1.77$ $\mathrm{eV\ \AA^{2}}$, $D_{2}=10.82$ $\mathrm{eV\ \AA^{2}}$, $A_{1}=0.30$ $\mathrm{eV\ \AA}$, $A_{2}=1.76$ $\mathrm{eV\ \AA}$, $B_{1}=2.55$ $\mathrm{eV\ \AA^{2}}$, $B_{2}=14.20$ $\mathrm{eV\ \AA^{2}}$, $M=-0.09$ eV, $m_{5}=3.8$ meV. The fitted energy dispersions along $k_{z}$, $k_{x}$, and $k_{101}$ directions, are presented in Figs. S2(a)-(c), where the red dashed (blue solid) lines are obtained from $k\cdot p$ model (\emph{ab initio} calculations). Note that there is some ambiguity in determining the parameters by fitting to the energy dispersion. Nevertheless, they suffice to describe the qualitative features of the band structure around $\Gamma$ point, especially the band inversion.

\subsection{Surface LDOS}
In Fig. S3(a), by constructing maximally localized Wannier functions on the basis of first-principles calculations~\cite{Marzari1997}, we present the local density of states of the (111) (or $z$-direction) surface terminated by a $\mathrm{MnBi_{2}Te_{4}}$ septuple layer. The Fermi surfaces of the massive surface state near and away from the conduction band bottom are shown in Figs. S3(b) and S3(c), with circular and triangular shapes, respectively. Both of them exhibit a spin-momentum locked helical spin-texture.
\begin{figure*}
  \centering
  \includegraphics[width=5.8in]{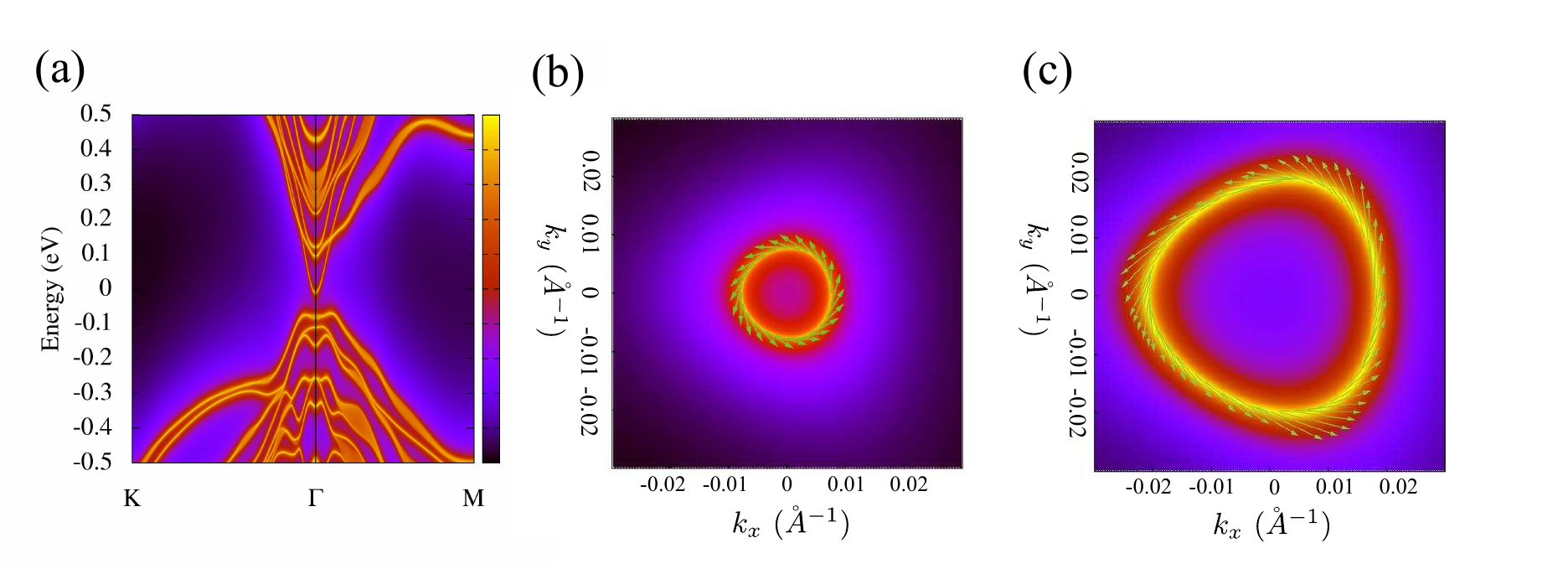}\\
  \caption{(a) The local density of states on the (111) surface terminated by the $\mathrm{Mn_{1}Bi_{2}Te_{4}}$ septuple layer. Fermi surfaces of the massive Dirac surface state at different energy level (b) near and (c) away from the conduction band minimum, with circular and triangular shapes, respectively. Both of them exhibit helical spin-momentum locked features.}\label{figs3}
\end{figure*}
\section{Calculation of the static axion angle}
As shown above, because of the $\mathcal{PT}$-symmetry, only the identity matrix $\mathbb{I}_{4\times 4}$ and the five linearly independent anticommuting Dirac matrices $\Gamma^{1,2,...,5}$ appear in the total Hamiltonian. The $\theta$ value of our model can thus be simply calculated via the formula \cite{li2010dynamical}
\begin{equation}
\label{thetacalc}
\theta=\frac{1}{4\pi}\int d^{3}k\frac{2|d|+d_{4}}{(|d|+d_{4})^{2}|d|^{3}}\epsilon^{ijkl}d_{i}\partial_{x}d_{j}\partial_{y}d_{k}\partial_{z}d_{l},\tag{S55}
\end{equation}
where $i,j,k,l$ take values from 1, 2, 3, 5 and $|d|=\sqrt{\sum_{n=1}^{5}d_{n}^{2}}$ with the lattice-regularized components of $d_{1,2,...,5}$ in the continuum model.

\section{Band structures for other compounds}
In Fig. S4, we present the detailed band structures obtained from first-principles calculations without and with SOC for other compounds, $X_{2}Y_{6}Z_{11}$ ($X=\mathrm{Eu,Mn}$, $Y=\mathrm{Bi,Sb}$, and $Z=\mathrm{Se, Te}$) in the $\mathrm{Mn}_{2}\mathrm{Bi}_{6}\mathrm{Te}_{11}$-family. The fatband structures have been labeled for each compound, which clearly demonstrate the SOC-induced band inversion around the $\Gamma$ point. This indicates that these compounds are also promising candidates for TI-based dynamical axion states. Figure S5 provides the insets of the energy gaps as a function of SOC strength near the minimum point in Fig. 4(b) in the main text.

\begin{figure*}
  \centering
  \includegraphics[width=7in]{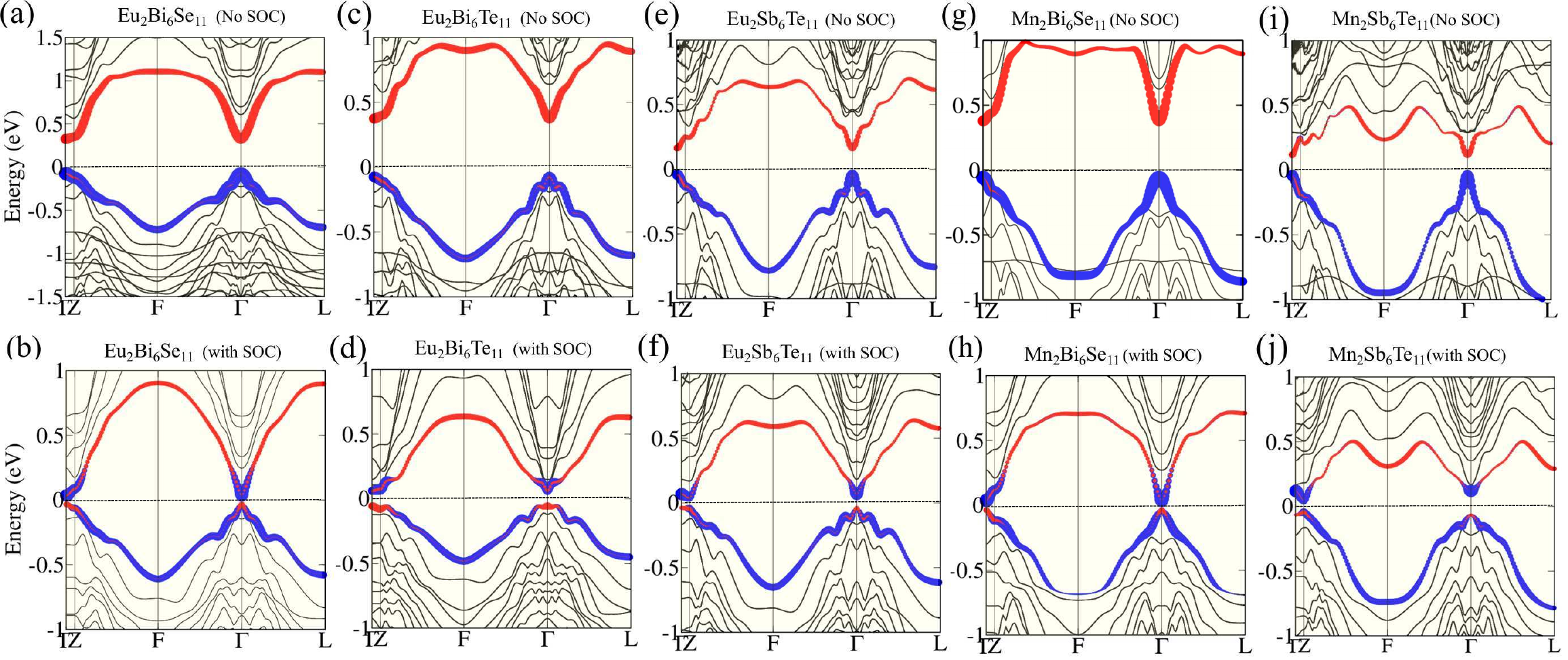}\\
  \caption{Band structures without (upper row) and with (lower row) SOC for other compounds in the $\mathrm{Mn}_{2}\mathrm{Bi}_{6}\mathrm{Te}_{11}$-family, namely, (from left to right), $\mathrm{Eu}_{2}\mathrm{Bi}_{6}\mathrm{Se}_{11}$, $\mathrm{Eu}_{2}\mathrm{Bi}_{6}\mathrm{Te}_{11}$, $\mathrm{Eu}_{2}\mathrm{Sb}_{6}\mathrm{Te}_{11}$, $\mathrm{Mn}_{2}\mathrm{Bi}_{6}\mathrm{Se}_{11}$, and, $\mathrm{Mn}_{2}\mathrm{Sb}_{6}\mathrm{Te}_{11}$. }\label{figs4}
\end{figure*}

\begin{figure*}
  \centering
  \includegraphics[width=6in]{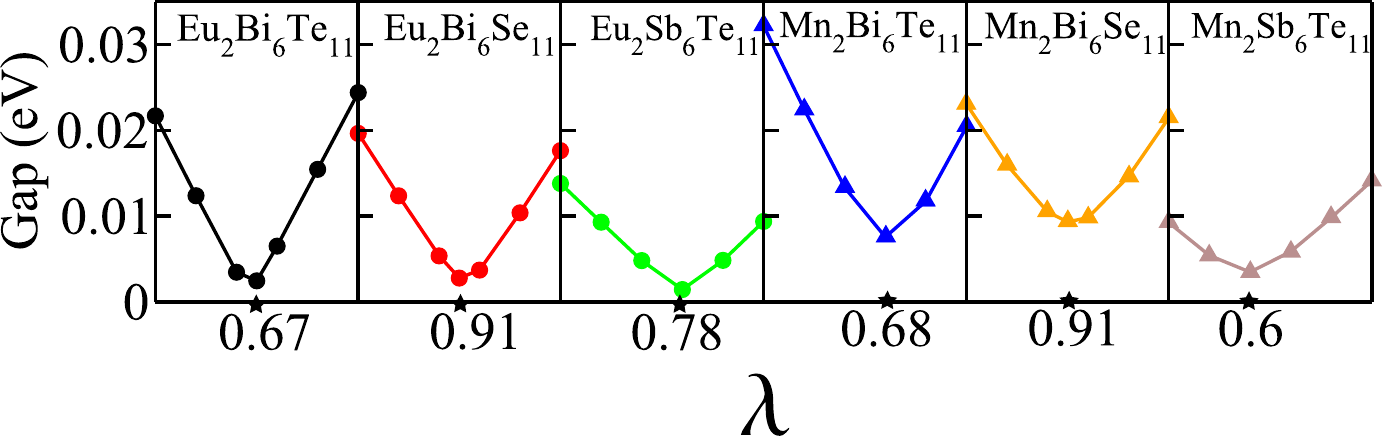}\\
  \caption{Insets of the energy gaps as a function of SOC strength near the minimum point in Fig. 4(b) in the main text.}\label{figs5}
\end{figure*}
\bibliography{SMRef}